\documentclass[reprint, aps, superscriptaddress, twocolumn]{revtex4-2}

\usepackage{graphicx,amsmath,amssymb,bm,booktabs}
\usepackage[T1]{fontenc}
\usepackage[utf8]{inputenc}
\usepackage{cmap}
\input{glyphtounicode}
\usepackage{xcolor}
\usepackage{orcidlink} 
\usepackage{tikz} 

\usetikzlibrary{arrows.meta,positioning,fit,calc} 
\hypersetup{
  colorlinks=true,
  linkcolor=blue,
  citecolor=blue,
  urlcolor=blue
}
\begin{document}

\title{Variational neural-network solution of the two-body $^{17}\mathrm{F}$ proton-halo problem with a Coulomb--Whittaker tail}

\author{Lucas A.~Souza\,\orcidlink{0000-0002-8711-956X}}
\email{lasouza@if.usp.br}
\affiliation{Minas Gerais, Brazil}
\author{Tobias Frederico\,\orcidlink{0000-0002-5497-5490}}
\email{tobias@ita.br}
\affiliation{Instituto Tecnol\'ogico de Aeron\'autica, 12.228-900, S\~ao Jos\'e dos Campos, SP, Brazil}

\begin{abstract}
We present a variational artificial neural-network (VANN) solution of
$^{17}$F in a two-body $^{16}$O$+p$ potential model. The calculation uses
a standard interaction from the literature as a controlled benchmark for
testing whether a neural variational ansatz can reproduce not only bound-state
energies and interior wave functions, but also the Coulomb--Whittaker tails
that control halo and peripheral-capture observables. The reduced radial wave
function is obtained by minimizing the Rayleigh quotient of the radial
Schr\"odinger Hamiltonian with the constraints required by each partial wave.
Because the variational energy can converge before the asymptotic
normalization is correct, the ansatz combines a neural interior with the
charged-particle Coulomb--Whittaker form. In the $s_{1/2}$ channel, the
Pauli-forbidden $0s_{1/2}$ component is computed and the physical one-node
branch is checked independently for forbidden-state contamination. The Coulomb--Whittaker-constrained VANN reproduces independent Numerov benchmarks for
the compact $d_{5/2}$ ground state and the extended $s_{1/2}$ halo state in
energy, nodes, rms radius, and overlap. The compact-state ANC agrees to within
one percent, while the halo ANC differs by about $4.2\%$, within the larger
numerical sensitivity of the asymptotic extraction. The continuum scattering states
are obtained by standard Numerov integration with Coulomb matching; only the
bound states are represented by the neural ansatz. Combined with these
$p$-wave scattering states, the VANN bound states yield astrophysical
$S$ factors consistent with published benchmarks and data within the accuracy
of the adopted two-body model. The results demonstrate the usefulness of
physically constrained neural wave functions for tail-sensitive nuclear
calculations and identify the asymptotic region as the most sensitive part of
the calculation.
\end{abstract}

\maketitle

\section{Introduction}
\label{sec:intro}
Direct radiative capture at low energy is controlled by the overlap
between a continuum entrance channel and the tail of the final bound
state. The reaction $^{16}$O$(p,\gamma)^{17}$F is a useful benchmark for
this situation because it combines a simple two-body structure with
observables that are highly sensitive to the asymptotic normalization of
the bound-state wave function: $S$ factors, elastic phase shifts, and
ANCs~\cite{Rolfs1973,Chow1975,Iliadis2008,Mohr2012,Tursunov2024,
Iliadis2022}. 

The $^{17}$F spectrum contains a compact
$J^\pi=5/2^+$ ground state and a weakly bound $J^\pi=1/2^+$ first excited
state. The latter has an extended $s_{1/2}$ proton tail and is commonly
discussed as a proton-halo configuration, as indicated by beta-decay
studies and microscopic structure calculations~\cite{Borge1993,Hagen2010}.
In this setting, reproducing the binding energy is not a sufficient test
of a numerical wave function: peripheral matrix elements depend directly
on the large-distance Coulomb--Whittaker tail and on the ANC.
This makes $^{17}$F a stringent test problem for neural wave-function
representations. A radial two-body Hamiltonian provides a transparent and
well-controlled setting, while the halo state and the capture observable
probe precisely the part of the wave function that is hardest to learn
from a variational energy alone. 
 
Two-body $^{16}$O+$p$ potential models remain attractive because the same
proton--core interaction can be used for bound states, elastic channels,
ANCs, and electromagnetic matrix elements.  Complementary descriptions of
$^{16}$O$(p,\gamma)^{17}$F include shell-model calculations~\cite{Bennaceur2000},
effective-field-theory treatments of proton halos and external capture
amplitudes~\cite{Ryberg2013}, and modern reaction-rate or Bayesian
analyses in which ANC information plays a central role~\cite{Iliadis2008,
Iliadis2022}.  These studies reinforce a practical lesson for any
numerical method applied to this reaction: the asymptotic normalization is
not a harmless detail of the wave function, but a quantity to be checked
explicitly.

Neural-network wave-function representations have become a useful tool in
scientific computing when they are treated as constrained variational
objects rather than as black-box interpolants~\cite{LeCun2015,
raissi2019pinns,karniadakis2021piml}. In nuclear and few-body physics,
related developments include neural-network quantum states~\cite{carleo2017nqs},
machine-learning studies of the deuteron~\cite{KeebleRios2020,Rozalen2024}, neural
solutions of few-body bound-state problems~\cite{LiLuoSunOrtega2026},
physics-informed treatments of Woods--Saxon and Dirac equations~\cite{Du2026},
and PINN-based approaches to scattering~\cite{Lei2026}. The present work
uses a controlled radial problem to isolate a specific issue that will
also appear in more complex applications: a variational neural ansatz may
learn the energetically important interior before it learns the physically
important tail.

The VANN used here represents the reduced radial wave function. Its loss
is based on the Rayleigh quotient of the radial Hamiltonian, with
auxiliary terms enforcing the constraints required by the channel. A
hybrid ansatz is adopted in which a multilayer perceptron describes the
nuclear interior and a Coulomb--Whittaker factor describes the
charged-particle tail. This construction is motivated by a negative
control: a flexible interior ansatz with a simple exponential envelope can
converge in energy and overlap while still failing in the far tail. Since
the low-energy capture amplitude is peripheral, this failure is physically
significant even when the wave functions appear nearly identical on a
linear scale.

There is one additional constraint in the $s_{1/2}$ channel. The
proton+$^{16}$O cluster Hamiltonian contains a deeply bound
Pauli-forbidden $0s_{1/2}$ configuration. A Rayleigh minimization in the
full $s$-wave space correctly finds that mathematical ground state, not
the physical halo state. The forbidden component must therefore be identified,
and orthogonality is enforced through an overlap penalty that selects the
physical one-node branch, as described in
Sec.~\ref{sec:training_protocol}.

The methodological contribution is to impose the Coulomb--Whittaker form,
control its matching to a neural interior, identify the Pauli-forbidden
component in the $s$-wave channel, enforce orthogonality through an overlap
penalty that selects the physical one-node branch, and propagate the residual tail error to
a peripheral capture observable. The exponential-tail control illustrates
the standard variational fact that accurate energies and global overlaps do
not by themselves guarantee accurate peripheral observables.

The paper is organized as follows.  Section~\ref{sec:method} defines the
two-body Hamiltonian, the VANN ansatz, the Coulomb--Whittaker tail, and
the treatment of the Pauli-forbidden $s_{1/2}$ component.  Section~\ref{sec:bound}
compares the VANN bound states with independent Numerov solutions.
Section~\ref{sec:scattering} describes the Coulomb-matched continuum
states used in the capture calculation.  Section~\ref{sec:capture}
presents the tail-sensitivity test and the resulting astrophysical
$S$ factors.  Section~\ref{sec:conclusions} summarizes the scope and
limitations of the calculation.

\section{Methodology: two-body physics and variational neural-network framework}
\label{sec:method}

\subsection{Two-body Hamiltonian}
\label{sec:hamiltonian}

In the two-body cluster approximation, $^{17}$F is described as a
valence proton moving relative to an inert $^{16}$O core. The reduced
radial wave function $u_{\ell j}(r)$ satisfies
\begin{equation}
\left[-\frac{\hbar^2}{2\mu}\frac{d^2}{dr^2}
+\frac{\hbar^2\ell(\ell+1)}{2\mu r^2}+V_{\ell j}(r)\right]
u_{\ell j}(r)=E u_{\ell j}(r),
\label{eq:radial}
\end{equation}
with $\int_0^\infty |u(r)|^2dr=1$. The proton--core reduced mass is
$\mu=m_p m_c/(m_p+m_c)$. The effective interaction is written as
\begin{equation}
V_{\ell j}(r)=V_N(r)+V_C(r)
+V_{SO}(r)\,\langle\bm{\ell}\!\cdot\!\mathbf{S}\rangle_{\ell j},
\label{eq:potential}
\end{equation}
where $V_N$ is a Woods--Saxon central interaction, $V_C$ is the
Coulomb potential of a uniformly charged sphere, and $V_{SO}$ is the
spin--orbit term. We use the V$_{\mathrm{M2}}$ parameter set of
Ref.~\cite{Tursunov2024} as the fixed Hamiltonian for the
$d_{5/2}$ ground state and the weakly bound $s_{1/2}$ halo state.

An independent Numerov calculation is performed with the same
Hamiltonian. The Numerov solution satisfies the regular condition
$u(r)\propto r^{\ell+1}$ near the origin and is matched in the external
region to the Coulomb--Whittaker form
\begin{equation}
u(r)\xrightarrow{r\rightarrow\infty}
C_{\ell j}W_{-\eta,\ell+1/2}(2\kappa r),
\qquad
\kappa=\frac{\sqrt{2\mu S_p}}{\hbar}.
\label{eq:numerov_tail}
\end{equation}
This calculation provides reference energies, wave functions, radii, and
ANCs. It is used only for validation, not as a training target for the
VANN.

\subsection{Hybrid neural ansatz and Coulomb--Whittaker tail}
\label{sec:ansatz}

The VANN represents the reduced radial wave function with a hybrid
ansatz,
\begin{equation}
u_\theta(r)=\left[1-B(r)\right]u_{\mathrm{int}}(r)
+B(r)u_{\mathrm{tail}}(r),
\label{eq:hybrid_ansatz}
\end{equation}
where the smooth switching function is
\begin{equation}
B(r)=\frac{1}{2}\left[1+\tanh\left(\frac{r-R_m}{a}\right)\right].
\label{eq:blend}
\end{equation}
The interior component is
\begin{equation}
u_{\mathrm{int}}(r)
=
r^{\ell+1}e^{-\alpha r}\,
\operatorname{MLP}_{\theta}\!\left[\log(1+r)\right],
\label{eq:interior}
\end{equation}
where ${\mathrm{MLP}}_\theta$ has two hidden layers of 64 neurons with
$\tanh$ activation. The decay parameter $\alpha>0$ and the MLP weights
are jointly trainable.

The tail component is
\begin{equation}
u_{\mathrm{tail}}(r)=C_\theta
\frac{W_{-\eta_\theta,\ell+1/2}(2\kappa_\theta r)}
{W_{-\eta_\theta,\ell+1/2}(2\kappa_\theta R_m)},
\label{eq:tail}
\end{equation}
with trainable amplitude $C_\theta$ (this is the normalized tail-basis amplitude, not identical to the physical ANC reported in Table~\ref{tab:bound_summary}). The inverse length
$\kappa_\theta$ is tied to the current Rayleigh energy of the trial
state,
\begin{equation}
\kappa_\theta =
\frac{\sqrt{2\mu |E_\theta|}}{\hbar},
\qquad E_\theta < 0 ,
\label{eq:kappa}
\end{equation}
and the Sommerfeld parameter is
\begin{equation}
\eta_\theta =
\frac{\mu Z_1Z_2 e^2}{\hbar^2\kappa_\theta}.
\label{eq:eta}
\end{equation}
A positive initial value of $\kappa_\theta$ defines the Whittaker tail before
the randomly initialized trial state becomes bound. Whenever the Rayleigh
energy is negative, the binding momentum is updated every 500 epochs using
Eq.~\eqref{eq:kappa}; $\eta_\theta$ and the Whittaker tail are then
reconstructed, and the updated tail is used in the next training block. Thus
$\kappa_\theta$ is determined from the variational energy and is not trained
as an independent network parameter.

For a charged bound state, the asymptotic behavior is not a free
exponential but a Coulomb--Whittaker function. This distinction is
important because the Rayleigh quotient is dominated by the region in
which the wave function has appreciable norm; it may become nearly
stationary before the small large-distance amplitude is correct. The
Whittaker component makes the physical tail part of the ansatz rather
than an after-the-fact diagnostic.

\subsection{Rayleigh variational objective}
\label{sec:loss}

For a trial function $u_\theta$, the energy minimized by the network is
the Rayleigh quotient
\begin{equation}
E_\theta=
\frac{\langle u_\theta|H|u_\theta\rangle}
{\langle u_\theta|u_\theta\rangle}.
\label{eq:rayleigh}
\end{equation}
On the radial grid this is evaluated as
\begin{equation}
E_\theta=
\frac{
\displaystyle
\frac{\hbar^2}{2\mu}\int |u_\theta'(r)|^2dr
+\int |u_\theta(r)|^2V_{\mathrm{eff}}(r)dr}
{\displaystyle\int |u_\theta(r)|^2dr}.
\label{eq:rayleigh_grid}
\end{equation}
The variational objectives used in the two channels are
\begin{equation}
 {\cal L}_{d_{5/2}}=E[u]+\lambda_{\rm bc}|u(r_{\max})|^2 ,
\label{eq:loss_d52}
\end{equation}
and
\begin{equation}
 {\cal L}_{s_{1/2}}=E[u]+\lambda_{\rm bc}|u(r_{\max})|^2
 +\lambda_P|\langle u|u_f\rangle|^2,
 \qquad \lambda_P=100 .
\label{eq:loss_s12}
\end{equation}
Normalization is imposed explicitly before evaluating $E[u]$ and is not a
separate penalty. Since $|u|^2$ has units fm$^{-1}$, the calculations use
$\lambda_{\rm bc}=1$~MeV~fm and $\lambda_P=100$~MeV; thus every term in
Eqs.~(\ref{eq:loss_d52})--(\ref{eq:loss_s12}) has units of MeV. No new
nondimensionalization is introduced. Value- and derivative-matching penalties were
implemented as diagnostic options but were not used in the calculations
reported here. No explicit projection of the trial wave function is applied
in the calculations reported here. No
target wave function, target ANC, target radius, target node position,
or empirical $S$-factor normalization is used during training. The
optimized outputs are the variational energy $E_\theta^*$ and the
corresponding wave function $u^*_{\mathrm{VANN}}(r)$.

\subsection{Channel constraints and training protocol}
\label{sec:training_protocol}

The same hybrid ansatz is used in the $d_{5/2}$ and $s_{1/2}$ channels,
but the variational spaces differ. In the $d_{5/2}$ channel, no
Pauli-forbidden $d$-wave state competes with the physical solution, and
the Rayleigh functional is minimized directly.

In the $s_{1/2}$ channel, the V$_{\mathrm{M2}}$ Hamiltonian supports a deeply
bound $0s_{1/2}$ configuration associated with an orbital already
occupied in the $^{16}$O core. We first obtain this forbidden component by
an unrestricted VANN minimization. Its normalized wave function
$u_f^{\rm VANN}$ is then kept fixed while the halo calculation minimizes
the Rayleigh objective with the additional Pauli-overlap penalty
$\lambda_P|\langle u|u_f^{\rm VANN}\rangle|^2$, which suppresses
the forbidden component and selects the physical one-node branch, in accord
with the orthogonality-condition treatment of Pauli-forbidden cluster
states~\cite{Saito1969}. An independent finite-difference diagonalization is
used only to validate the forbidden-state energy and wave function. The
optimized halo is nearly orthogonal to the fixed VANN reference,
$|\langle u_{s_{1/2}}^{\rm VANN}|u_f^{\rm VANN}\rangle|=1.61\times10^{-3}$;
its overlap with the independent FD forbidden state is $5.36\times10^{-3}$.

\begin{table}[ht]
\centering
\caption{Neural-network architecture and training parameters.}
\label{tab:vann_params}
\scriptsize
\setlength{\tabcolsep}{4pt}
\begin{tabular}{lcc}
\toprule
Parameter & $d_{5/2}$ & $s_{1/2}$ \\
\midrule
\multicolumn{3}{c}{Architecture} \\
\midrule
Hidden layers $\times$ neurons & \multicolumn{2}{c}{$2\times64$} \\
Activation & \multicolumn{2}{c}{$\tanh$} \\
Input & \multicolumn{2}{c}{$\ln(1+r)$, $r_0=1$~fm} \\
\midrule
\multicolumn{3}{c}{Hybrid ansatz} \\
\midrule
$R_m$ (fm) & 12 & 20 \\
$a$ (fm) & \multicolumn{2}{c}{2} \\
\midrule
\multicolumn{3}{c}{Grid} \\
\midrule
$r_{\mathrm{min}}$ (fm) & \multicolumn{2}{c}{$10^{-5}$} \\
$r_{\mathrm{max}}$ (fm) & 100 & 200 \\
$N$ points & 5,000 & 10,000 \\
\midrule
\multicolumn{3}{c}{Training} \\
\midrule
Optimizer, lr & \multicolumn{2}{c}{Adam, $5\times10^{-4}$} \\
Random seed & \multicolumn{2}{c}{1234} \\
Epochs & \multicolumn{2}{c}{20,000} \\
\midrule
\multicolumn{3}{c}{Loss weights} \\
\midrule
$\lambda_E$ (Rayleigh) & \multicolumn{2}{c}{1 (dimensionless)} \\
$\lambda_{\mathrm{bc}}$ (boundary) & \multicolumn{2}{c}{1 MeV fm} \\
$\lambda_P$ (Pauli overlap) & N/A & 100 MeV \\
\midrule
\multicolumn{3}{c}{Whittaker tail (energy-updated converged values)} \\
\midrule
$\kappa$ (fm$^{-1}$) & 0.164794 & 0.069859 \\
$\eta$ & 1.58465 & 3.73811 \\
\midrule
\multicolumn{3}{c}{Final values} \\
\midrule
$\alpha$ (fm$^{-1}$) & 0.5981 & 0.1367 \\
$C_{\mathrm{tail}}$ & 0.6079 & 0.2010 \\
$E_\theta^*$ (MeV) & $-0.598989$ & $-0.107642$ \\
$\Delta E$ (keV) & 0.061 & 1.682 \\
\bottomrule
\end{tabular}
\end{table}

The calculations reported here used the numerical parameters listed in
Table~\ref{tab:vann_params}. To check that the conclusions were not tied
to a single setup, we varied the grid size by $\pm2000$ points, the
matching radius by $R_m\pm2$~fm, the blending width by $a\pm1$~fm, and
the random seed. These changes shifted the bound-state energies only at
the few-keV level and kept the overlaps with the Numerov benchmarks
above $0.999$. The ANC, especially for the halo state, was the most
sensitive quantity, with variations up to $\sim10$~fm$^{-1/2}$.

\section{Bound states}
\label{sec:bound}

We now apply the VANN framework to the two bound states of
$^{17}$F in the V$_{\mathrm{M2}}$ two-body model.  The two channels provide
complementary tests of the ansatz.  The compact $d_{5/2}$ ground state
tests whether Rayleigh minimization with an embedded
Coulomb--Whittaker tail can reproduce a conventional bound state and its
asymptotic normalization.  The weakly bound $s_{1/2}$ state is more
demanding: the physical halo must be isolated from a deeply bound
Pauli-forbidden configuration, and its long Coulomb tail must be
stabilized.

\subsection{Ground state: $d_{5/2}$}
\label{sec:d52}

The $d_{5/2}$ ($J^\pi=5/2^+$) ground state is the cleanest bound-state
test because no Pauli-forbidden $d$-wave configuration competes with the
physical solution.  It is more compact than the halo state, but its
asymptotic normalization is still relevant for radiative capture: the
E1 matrix element samples radii beyond the nuclear surface, where the
bound-state tail controls the peripheral contribution.

\begin{figure*}[htpb]
\centering
\begin{tikzpicture}[
font=\small,
>=Latex,
node distance=10mm and 8mm,
box/.style={draw,rounded corners,align=center,minimum height=6mm,text width=3.1cm,inner sep=1mm},
widebox/.style={draw,rounded corners,align=center,minimum height=7mm,text width=2.8cm,inner sep=1mm},
arr/.style={-Latex,line width=0.8pt}
]
\node[box,text width=1.7cm,minimum height=5mm,font=\small] (r) {radial grid \\$r_i$};
\node[box, text width=3cm,right=of r] (ansatz) {ansatz\\ $u_\theta\rightarrow  u_{\mathrm{int}} + u_{\mathrm{tail}}$};
\node[box, text width=2.6cm ,right=of ansatz] (mlp) {MLP interior \\ $[1,64,64,1]$ };
\node[box, text width=3cm,right=of mlp] (loss) {Rayleigh loss\\ $E_\theta=\frac{\langle u_\theta|H|u_\theta\rangle}{\langle u_\theta|u_\theta\rangle}$};
\node[box, text width=1.7cm,right=of loss] (out) {output\\ $E_\theta^*$\\ $u^*_{\mathrm{VANN}}(r)$};
\node[box,text width=3.7cm,below=6mm of ansatz] (tail) {Coulomb tail\\ $u_{\mathrm{tail}}\propto W_{-\eta,5/2}(2\kappa r)$};
\draw[arr] (r) -- (ansatz);
\draw[arr] (ansatz) -- (mlp);
\draw[arr] (mlp) -- (loss);
\draw[arr] (loss) -- (out);
\draw[arr] (tail.north) -- (ansatz.south);
\node[draw,dashed,rounded corners,fit=(r)(ansatz)(mlp)(loss)(out)(tail),inner sep=1mm,
label=above:{\textbf{$d_{5/2}$ variational workflow}}] {};
\end{tikzpicture}%
\caption{Workflow used to obtain the $d_{5/2}$ bound-state solution.  The
radial grid is mapped to the hybrid VANN ansatz, whose interior is
represented by a neural network and whose large-distance behavior is
fixed by the Coulomb--Whittaker form.  Minimization of the Rayleigh
functional gives the optimized energy and wave function.}
\label{fig:d52_workflow}
\end{figure*}
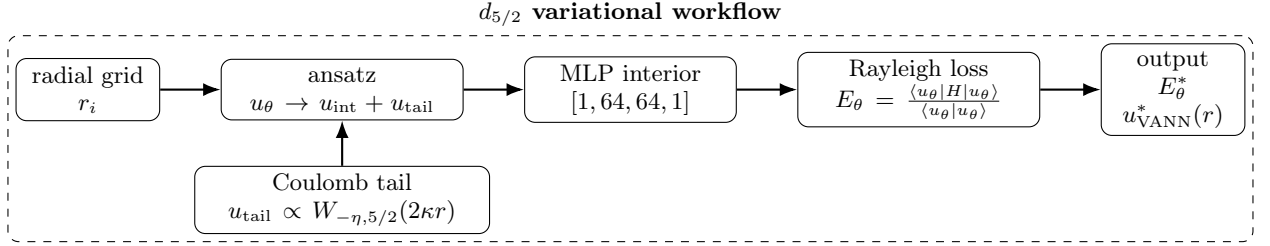

\begin{figure}[htpb]
\includegraphics[width=\columnwidth]{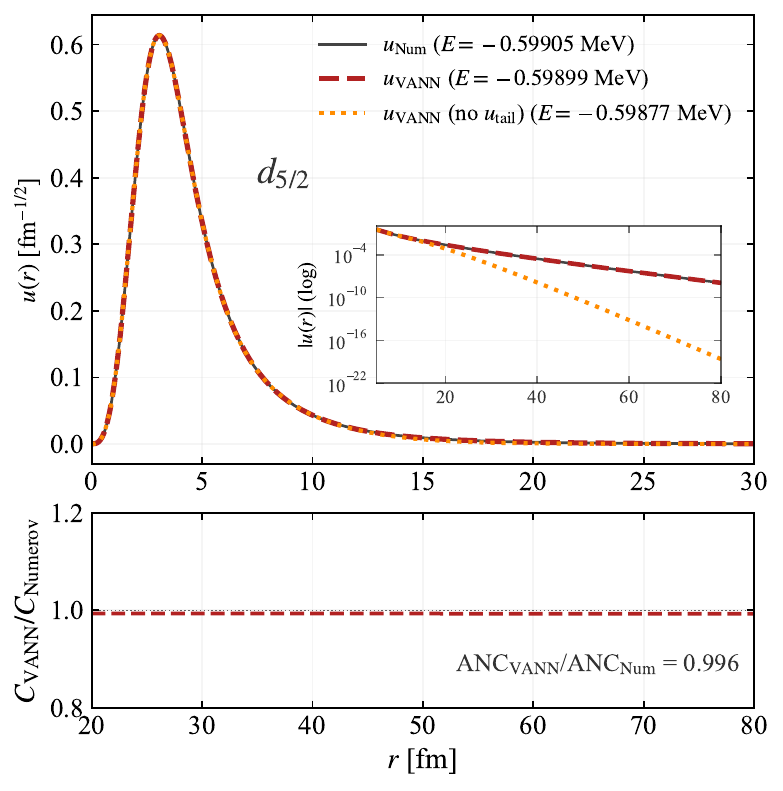}
\caption{Ground-state $d_{5/2}$ radial wave function obtained with the
VANN hybrid ansatz and compared with the independent Numerov benchmark.
The inset shows the logarithmic tail.  The Whittaker-constrained VANN
reproduces both the interior and the Coulomb asymptotic behavior.
The exponential-envelope control uses the same architecture, grid, and
training protocol, differing only in the replacement of the Whittaker
tail by a simple exponential decay; it matches the interior profile but
decays too rapidly at large radius. The lower panel shows the direct ratio
$u_{\rm VANN}(r)/u_{\rm Numerov}(r)$ (the common Numerov-energy Whittaker
factor used by the plotting script cancels), which remains close to unity
across the asymptotic region, confirming that the Whittaker-constrained
VANN reproduces the correct ANC for the compact ground state.}
\label{fig:d52_wave}
\end{figure}

Figure~\ref{fig:d52_workflow} summarizes the training workflow for the
$d_{5/2}$ channel.  The trained VANN solution is compared with an
independent Numerov solution of the same Hamiltonian in
Fig.~\ref{fig:d52_wave}.

The VANN and Numerov wave functions are nearly indistinguishable on the
linear scale.  The VANN energy is $E=-0.598989$~MeV, compared with the
Numerov value $E=-0.59905$~MeV, and the overlap is $1.00000$.  The rms
radius is $r_{\mathrm{rms}}=3.85$~fm, close to the Numerov value
$3.85$~fm.  The ANC extracted from the Whittaker ratio is
$C_{\mathrm{VANN}}=1.04$~fm$^{-1/2}$, matching the Numerov reference
$C_{\mathrm{Num}}=1.04$~fm$^{-1/2}$.

The exponential-envelope control in Fig.~\ref{fig:d52_wave} isolates
the role of the asymptotic constraint.  Both variants are compared
quantitatively in Table~\ref{tab:d52_control}.  The exponential control
reproduces the binding energy to within $0.28$~keV and achieves an
overlap of $0.99998$, yet its ANC is effectively zero
($C\approx0.01$~fm$^{-1/2}$) and its $S$ factor at $20$~keV is smaller
by about a factor of $8.6$ than the Whittaker-tail result.  This ablation
test is the strongest diagnostic of the calculation: a trial wave
function may pass standard bound-state checks while failing for a
tail-sensitive capture observable.

\begin{table}[h]
\centering
\caption{Diagnostic control for the $d_{5/2}$ ground state.
The exponential-envelope VANN matches the interior but fails in
tail-sensitive observables.}
\label{tab:d52_control}
\small
\begin{tabular}{lccc}
\toprule
& Whittaker & Exponential & Numerov \\
\midrule
$E$ (MeV) & $-0.598989$ & $-0.598774$ & $-0.59905$ \\
$r_{\mathrm{rms}}$ (fm) & 3.85 & 3.84 & 3.85 \\
Overlap & 1.00000 & 0.99998 & 1.00000 \\
ANC (fm$^{-1/2}$) & 1.04 & $\sim0.01$ & 1.04 \\
$S_0(20~{\mathrm{keV}})$ (keV$\cdot$b) & 0.386 & 0.0449 & 0.387 \\
\bottomrule
\end{tabular}
\end{table}

\subsection{Halo state: $s_{1/2}$}
\label{sec:s12}

The $s_{1/2}$ ($J^\pi=1/2^+$) state probes the most delicate part of the
calculation.  Its small separation energy produces a long
Coulomb--Whittaker tail, making the ANC and rms radius more sensitive to
the asymptotic region than in the compact ground state.  In addition,
the unconstrained $s$-wave Hamiltonian contains a deeply bound
Pauli-forbidden $0s_{1/2}$ state.

Before selecting the physical one-node branch, we minimize the unrestricted
$s$-wave Rayleigh functional with a VANN and obtain the deeply bound
Pauli-forbidden configuration. Its normalized wave function
$u_f^{\rm VANN}(r)$ is stored and kept fixed as the reference entering the
Pauli-overlap penalty. The normalized forbidden-state wave function obtained
from the unrestricted VANN calculation was subsequently evaluated with the
final Hamiltonian discretization, yielding $E_f^{\rm VANN}=-30.18844$~MeV.
An independent three-point finite-difference diagonalization gives
$E_f^{\rm FD}=-30.31534$~MeV, while the corresponding normalized wave functions
have an overlap of $0.999968$. The forbidden-state run used the same $2\times64$
architecture, activation, seed, optimizer, learning rate, and hybrid ansatz as
the halo calculation. The FD calculation thus
provides an external validation rather than a training target. The next FD bound eigenvalue is
$-0.10669$~MeV, independently confirming the physical one-node branch.

\begin{figure*}[htpb]
\centering
\begin{tikzpicture}[
font=\small,
>=Latex,
node distance=7mm and 6mm,
box/.style={draw,rounded corners,align=center,minimum height=7mm,text width=2.3cm,inner sep=1.mm},
arr/.style={-Latex,line width=1.0pt, shorten >=1pt,shorten <=1pt}
]
\node[box, text width=1.7cm] (r1) {radial grid\\$r_i$};
\node[box, text width=2.5cm,right=of r1] (ans1) {hybrid VANN\\ansatz};
\node[box, text width=1.8cm,right=of ans1] (net1) {\small MLP interior};
\node[box, text width=2.0cm,right=of net1] (loss1) {\small unrestricted\\Rayleigh loss};
\node[box, text width=3.5cm,right=of loss1] (out1) {{\small forbidden-state source}\\$E_f$\\$u_f(r)$};
\draw[arr] (r1) -- (ans1);
\draw[arr] (ans1) -- (net1);
\draw[arr] (net1) -- (loss1);
\draw[arr] (loss1) -- (out1);
\node[draw,dashed,rounded corners,fit=(r1)(ans1)(net1)(loss1)(out1),inner sep=1.mm,
label=above:{\textbf{(a) Unconstrained VANN forbidden branch}}] (branch1) {};

\node[box, text width=1.7cm,below=10mm of r1] (r2) {radial grid\\$r_i$};
\node[box, text width=2.5cm,right=of r2] (ans2) {ansatz\\ $u_\theta\rightarrow  u_{\mathrm{int}} + u_{\mathrm{tail}}$};
\node[box, text width=1.3cm,right=of ans2] (net2) {\small MLP interior};
\node[box, text width=3.4cm,right=of net2] (pauli) {Pauli-overlap penalty\\$\lambda_P|\langle u_f|u_\theta\rangle|^2$};
\node[box, text width=2cm,right=of pauli] (loss2) {\small Rayleigh loss\\one-node branch};
\node[box, text width=1.7cm,right=of loss2] (out2) {{\small halo output}\\$E^*_{\theta}$\\$u^*_{\mathrm{VANN}}(r)$};
\draw[arr] (r2) -- (ans2);
\draw[arr] (ans2) -- (net2);
\draw[arr] (net2) -- (pauli);
\draw[arr] (pauli) -- (loss2);
\draw[arr] (loss2) -- (out2);
\draw[->, thick] (out1.south) -- (pauli.north);
\node[draw,dashed,rounded corners,fit=(r2)(ans2)(net2)(pauli)(loss2)(out2),inner sep=1.mm,
label=below:{\textbf{(b) Pauli-penalized halo branch}}] {};
\end{tikzpicture}%
\caption{Penalty-only treatment in the $s_{1/2}$ channel. The deeply bound
forbidden configuration is first obtained by an unrestricted VANN
minimization and validated independently by finite differences. Its normalized
wave function is then kept fixed in the penalty
$\lambda_P|\langle u|u_f^{\rm VANN}\rangle|^2$, which strongly suppresses the
forbidden component and selects the physical weakly bound one-node branch.}
\label{fig:s12_workflow}
\end{figure*}
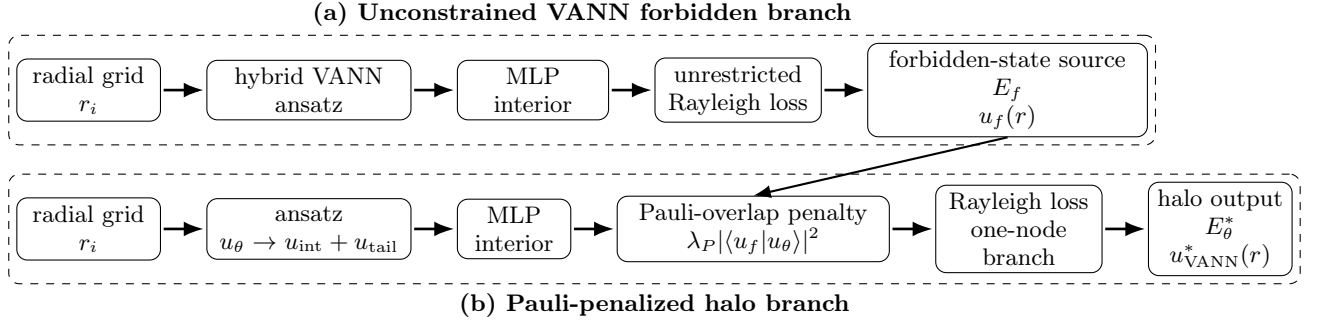

Figure~\ref{fig:s12_workflow} displays the two-step identification of the
forbidden state and the physical one-node $s_{1/2}$ branch.

\begin{figure}[htpb]
\includegraphics[width=\columnwidth]{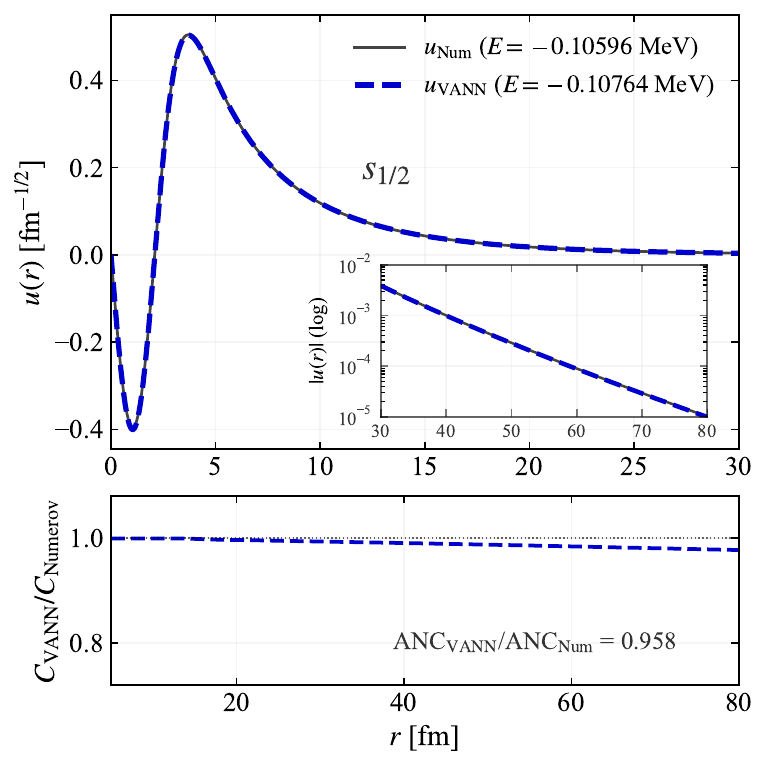}
\caption{Halo $s_{1/2}$ radial wave function obtained with the
one-node VANN ansatz and compared with the independent Numerov
benchmark.  The VANN solution is obtained without using the Numerov
wave function as a training target.  The inset highlights the long
Coulomb--Whittaker tail in logarithmic scale. The lower panel shows the
direct ratio $u_{\rm VANN}(r)/u_{\rm Numerov}(r)$; the systematic downward
drift reflects the slightly faster decay of the VANN tail, a
consequence of the small energy mismatch ($\Delta E\approx1.7$~keV)
between the VANN and Numerov solutions being amplified in the
asymptotic region.}
\label{fig:s12_wave}
\end{figure}

The one-node VANN yields $E=-0.107642$~MeV, $1.682$~keV away from
the Numerov benchmark for the same Hamiltonian.  The rms radius is
$r_{\mathrm{rms}}=5.17$~fm, the node number is one, and the overlap with the
Numerov halo wave function is $0.99999$.  The ANC is
$C_{\mathrm{VANN}}=70.90$~fm$^{-1/2}$, compared with
$C_{\mathrm{Num}}\simeq74$~fm$^{-1/2}$.  The residual $4.2\%$ difference
lies within the approximately $10$~fm$^{-1/2}$ sensitivity found under grid,
matching-window, and seed variations. For both methods,
$C(r)=|u(r)/W_{-\eta,1/2}(2\kappa r)|$ was averaged pointwise over
$20\le r\le60$~fm; each ANC used the Whittaker function constructed with
that method's own energy and hence its own $\kappa$ and $\eta$.

This behavior has the standard variational interpretation. The energy is
stationary and changes only at second order in a small wave-function error,
whereas the ANC and a peripheral transition matrix element respond at first
order. The tail also carries little norm, so a near-unit global overlap does
not tightly constrain its relative error. The $1.682$~keV binding-energy
difference increases $\kappa$ by about $0.79\%$; because the Whittaker tail
contains the exponential decay scale, matching equal amplitudes at 20~fm
predicts an ANC ratio $0.961$ and a finite-radius wave-function ratio $0.981$
at 80~fm. These values are close to the observed $0.958$ ANC ratio and the
lower panel of Fig.~\ref{fig:s12_wave}. The residual difference therefore
reflects first-order asymptotic sensitivity amplified by the small
binding-energy mismatch and should not be interpreted as a
neural-network-specific failure.

Figure~\ref{fig:s12_wave} shows the resulting halo wave function.  The
agreement with Numerov on the logarithmic tail is the key validation for
the subsequent capture calculation, because the external E1 amplitude is
controlled by the same large-distance region.

\subsection{Bound-state summary and convergence}
\label{sec:bound_summary}

\begin{table}[htpb]
\centering
\caption{Bound-state observables obtained with the VANN ansatz and
compared with the independent Numerov benchmark.  Energies are in MeV,
radii in fm, and ANCs in fm$^{-1/2}$.}
\label{tab:bound_summary}
\begin{tabular}{llccccc}
\toprule
State & Method & $E$ & $r_{\mathrm{rms}}$ & nodes & ANC & overlap \\
\midrule
$d_{5/2}$ & Numerov & $-0.59905$ & 3.85 & 0 & 1.04 & 1.00000 \\
$d_{5/2}$ & VANN    & $-0.598989$ & 3.85 & 0 & 1.04 & 1.00000 \\
$s_{1/2}$ & Numerov & $-0.10596$ & 5.17 & 1 & 74.0 & 1.00000 \\
$s_{1/2}$ & VANN    & $-0.107642$ & 5.17 & 1 & 70.90 & 0.99999 \\
\bottomrule
\end{tabular}
\end{table}

Table~\ref{tab:bound_summary} collects the final bound-state
observables.  The agreement in energy, node structure, radius, and
overlap shows that the VANN identifies the correct eigenstates of the
two-body Hamiltonian.  The ANC comparison is the stricter test: it is
essentially exact for the compact $d_{5/2}$ state and remains within
about $4.2\%$ for the extended $s_{1/2}$ halo.

\begin{figure}[htpb]
\includegraphics[width=\columnwidth]{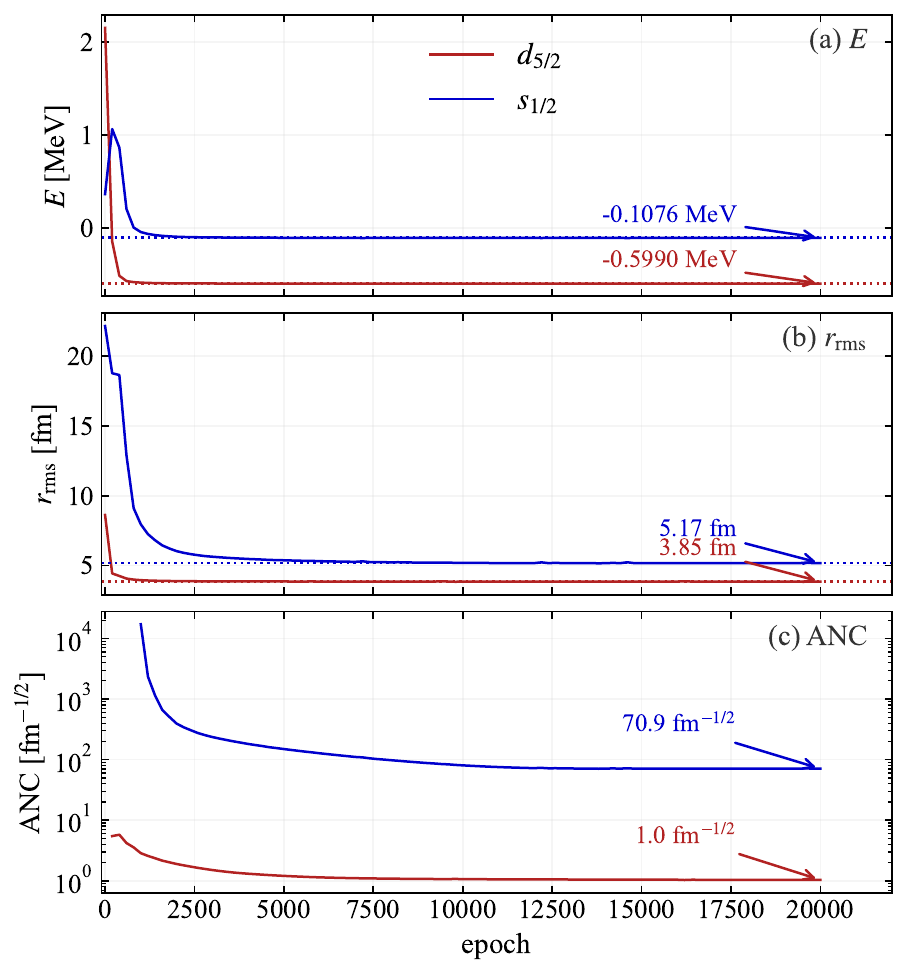}
\caption{Evolution of the Rayleigh energy, rms radius, and ANC during
VANN training for both bound states.  The energy converges first,
followed by the radius; the ANC requires the largest number of epochs,
particularly for the $s_{1/2}$ halo.  Dashed lines in panels (a) and
(b) mark the binding energies and rms radii obtained from the
V$_{\mathrm{M2}}$ potential calibration of
Ref.~\cite{Tursunov2024}, shown for orientation.}
\label{fig:convergence_physics}
\end{figure}

Figure~\ref{fig:convergence_physics} summarizes the training evolution
of the main physical observables.  The Rayleigh energy stabilizes before
the rms radius, while the ANC converges last, especially in the
$s_{1/2}$ halo channel.  This hierarchy confirms that the ANC is the slowest quantity to
stabilize during Rayleigh training, particularly for the
$s_{1/2}$ halo.
Grid size, matching radius, and random seed were varied around the
reported values in Table~\ref{tab:vann_params}. The bound-state
energies shifted by at most a few keV and the overlaps with Numerov
remained above $0.999$ in all cases. The ANC of the compact $d_{5/2}$
state was stable at $1.04$~fm$^{-1/2}$, while the $s_{1/2}$ halo ANC
varied by up to $\sim10$~fm$^{-1/2}$ depending on the grid resolution,
reflecting the greater tail sensitivity of the extended state.
An independent three-point finite-difference scan gives VANN--FD differences
of $-0.08$ to $+0.06$~keV for $d_{5/2}$ and $-1.56$ to $-0.95$~keV for
$s_{1/2}$ as $\Delta r$ is changed from 0.020 to 0.010~fm and
$r_{\max}$ from 100 to 200~fm. Their discretization dependence shows why
comparison of the two numerical schemes is not a formal test of the
variational upper-bound property.

\section{Continuum states}
\label{sec:scattering}

Having established the bound-state sector, we now construct the
positive-energy entrance-channel wave functions that enter the
radiative-capture matrix elements. These states describe the relative
motion of the incoming proton and the $^{16}$O core before photon
emission. For the $E1$ transitions considered below, the relevant
low-energy entrance waves are the $p_{1/2}$ and $p_{3/2}$ continua.

The continuum states are not obtained with the VANN ansatz. For each
chosen center-of-mass energy $E>0$, the energy is an input rather than
an eigenvalue. The radial Schrödinger equation is integrated outward
with the regular boundary condition
$u_{\ell j}(r)\propto r^{\ell+1}$ near the origin, and the resulting
solution is matched to the regular and irregular Coulomb functions in
the asymptotic region,
\begin{equation}
\begin{split}
u_{\ell j}^{(+)}(E,r)
\xrightarrow{r\rightarrow\infty}
N_E\Big[
&F_\ell(\eta,kr)\cos\delta_{\ell j}(E) \\
&+G_\ell(\eta,kr)\sin\delta_{\ell j}(E)
\Big],
\end{split}
\label{eq:scatt_asymptotic}
\end{equation}
where $F_\ell$ and $G_\ell$ are the regular and irregular Coulomb
functions, $k=\sqrt{2\mu E}/\hbar$, and
$\eta=\mu Z_1Z_2 e^2/(\hbar^2 k)$ is the Sommerfeld parameter.
The quantity $\delta_{\ell j}(E)$ is the nuclear phase shift relative
to the pure Coulomb solution. We use unit standing-wave asymptotic
amplitude: the outward-integrated Numerov solution is rescaled so that
$N_E=1$ in Eq.~\eqref{eq:scatt_asymptotic}. This is neither unit incoming
flux nor $\delta(E-E')$ normalization; the incident-velocity factor is
included explicitly in the absolute cross-section prefactor below.

The phase shift is extracted from the logarithmic derivative of the
Numerov solution at the matching radius:
\begin{equation}
L(E,R_m)
=
\left.
\frac{u'_{\ell j}(E,r)}{u_{\ell j}(E,r)}
\right|_{r=R_m}.
\label{eq:log_derivative}
\end{equation}
Denoting derivatives of the Coulomb functions with respect to
$\rho=kr$ by primes, the matching condition gives
\begin{equation}
\tan\delta_{\ell j}(E)
=
\frac{
kF'_\ell(\eta,kR_m)-L(E,R_m)F_\ell(\eta,kR_m)
}{
L(E,R_m)G_\ell(\eta,kR_m)-kG'_\ell(\eta,kR_m)
}.
\label{eq:phase_matching}
\end{equation}
  
\begin{figure}[htpb]
\includegraphics[width=\columnwidth]{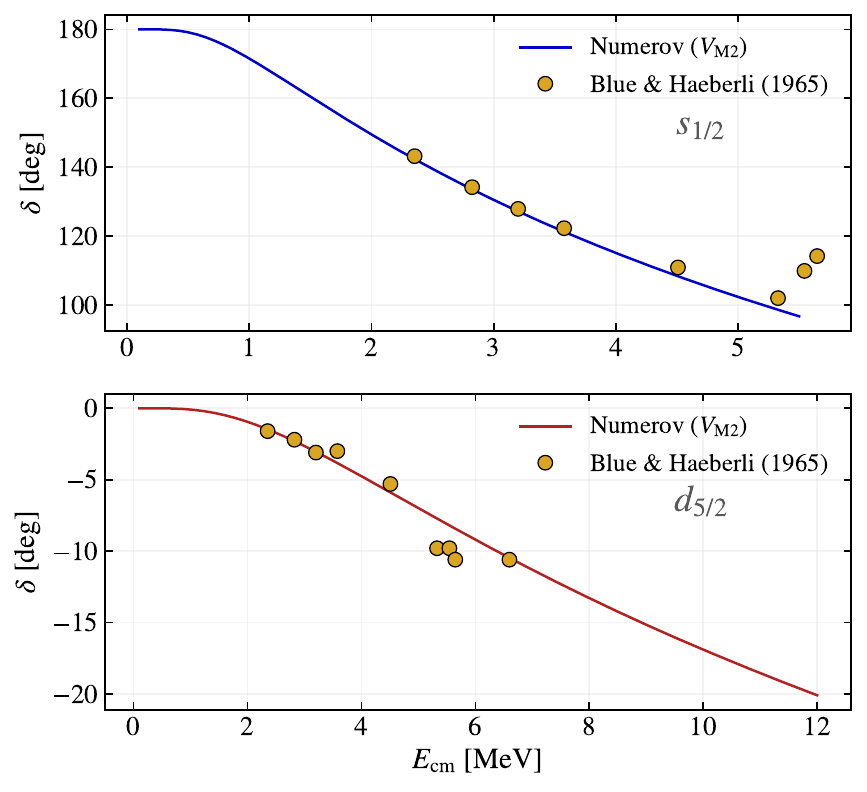}
\caption{Elastic phase shifts for $p+{}^{16}$O scattering in the
$^{2}S_{1/2}$ and $^{2}D_{5/2}$ partial waves,
computed with the same V$_{\mathrm{M2}}$ potential used for the bound states.
Solid curves are the Numerov calculation with Coulomb matching;
symbols are the experimental data of Blue and
Haeberli~\cite{Blue1965}.
The $^{2}S_{1/2}$ phases are shifted by $+180^\circ$ to display
the same branch as the calculated curve, consistent with
Levinson's theorem for the two $s$-wave bound states supported
by the potential.}
\label{fig:phase_shifts}
\end{figure}

Figure~\ref{fig:phase_shifts} complements the bound-state benchmarks by
testing the V$_{\mathrm{M2}}$ proton--core interaction in elastic
scattering channels.  The displayed $S$- and $D$-wave phase shifts are
not the $p$-wave entrance states themselves, but they correspond to the
same partial waves that support the two bound states studied in this
work.  They therefore provide an independent check that the adopted
interaction gives a reasonable description of the measured
phase-shift systematics within the two-body model.  The $E1$ capture
matrix elements are evaluated with the Coulomb-matched $p_{1/2}$ and
$p_{3/2}$ entrance waves described above.

\begin{figure}[htpb]
\includegraphics[width=\columnwidth]{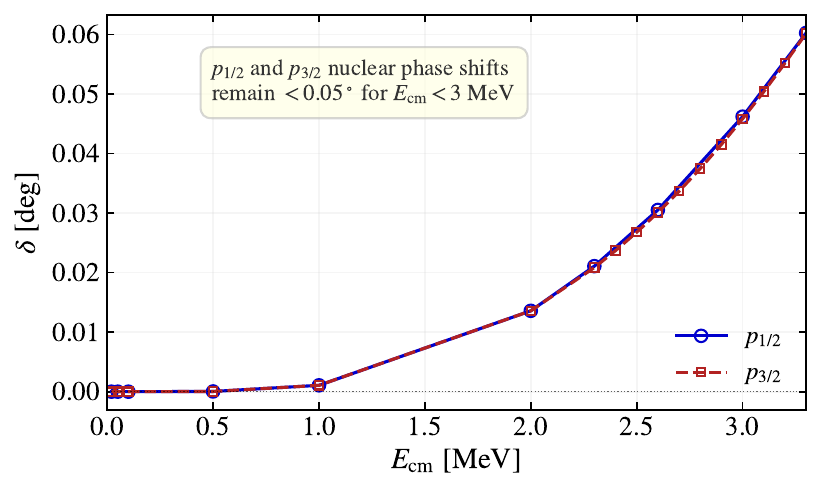}
\caption{Nuclear phase shifts for the $p_{1/2}$ and $p_{3/2}$ entrance
channels computed with the V$_{\mathrm{M2}}$ potential.
Both phase shifts remain below $0.05^\circ$ for
$E_{\mathrm{cm}}<3$~MeV, confirming that the $p$-wave continuum is
Coulomb-dominated in the energy range relevant for the capture
calculation.}
\label{fig:phase_shifts_pwave}
\end{figure}

Figure~\ref{fig:phase_shifts_pwave} complements the $S$- and $D$-wave
phase-shift check of Fig.~\ref{fig:phase_shifts} by directly validating
the entrance-channel partial waves used in the $E1$ matrix elements.
The nuclear component of the $p_{1/2}$ and $p_{3/2}$ scattering states
is negligible at low energy: for $E_{\mathrm{cm}}<1$~MeV the phase
shifts are below $10^{-3}$~degrees, and they remain under
$0.05^\circ$ up to $3$~MeV.  The entrance-channel continuum is
therefore essentially pure Coulomb in the capture region, and the
numerical construction of Sec.~\ref{sec:scattering} reduces to the
Coulomb-matched form of Eq.~\eqref{eq:scatt_asymptotic} with
$\delta_{\ell j}\approx0$.  This indicates the weak sensitivity expected
from the nuclear part of the $p$-wave interaction in the low-energy region
considered here: the capture is peripheral and
controlled by the bound-state tail and the Coulomb barrier
penetrability.

\begin{figure}[htpb]
\includegraphics[width=\columnwidth]{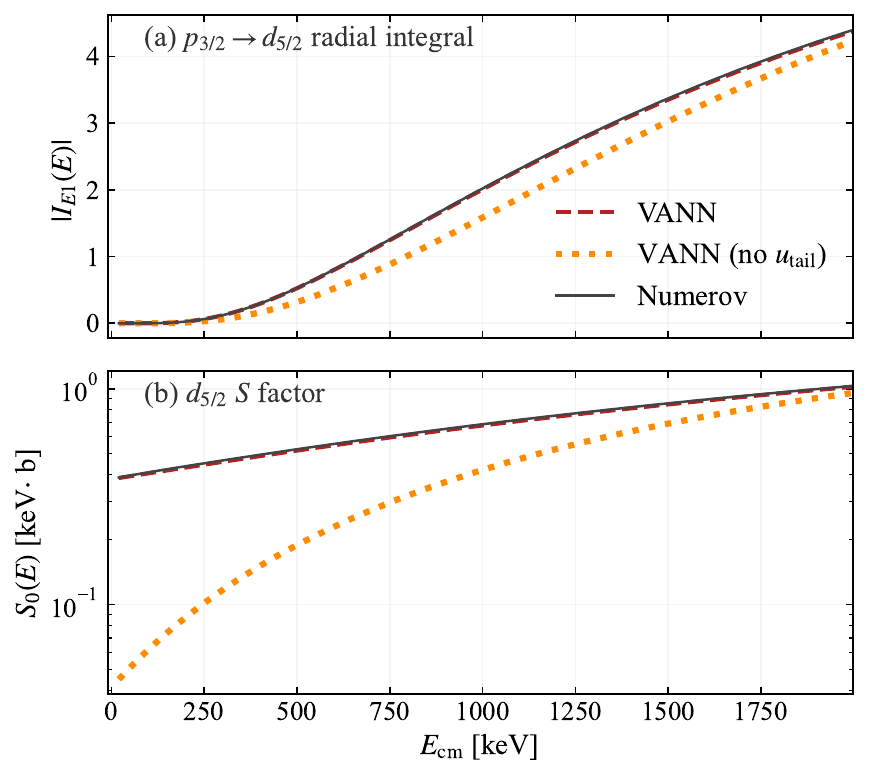}
\caption{Impact of the bound-state asymptotic behavior on the
$d_{5/2}$ capture matrix element.
Panel (a) compares the $E1$ radial integral for the
$p_{3/2}\to d_{5/2}$ transition obtained with the VANN bound state,
the VANN variant without the Whittaker tail, and the Numerov
benchmark.
Panel (b) shows the corresponding $S$ factor.
The $p_{1/2}\to d_{5/2}$ channel is $E1$-forbidden by angular-momentum
coupling (Table~\ref{tab:channel_factors}), so $p_{3/2}\to d_{5/2}$
is the only entrance channel contributing to ground-state capture,
making it the cleanest diagnostic for the tail-sensitivity test.}
\label{fig:d52_tail_impact}
\end{figure}

\section{Radiative capture and astrophysical $S$ factor}
\label{sec:capture}

The capture calculation follows the standard two-body direct-capture
framework.  The $E1$ operator for the core--proton relative coordinate is
$\hat{\mathcal O}_{E1}=e_{\mathrm{eff}}\,r\,Y_{1m}(\hat r)$, with the
effective charge
$e_{\mathrm{eff}}=e\,(Z_p m_c-Z_c m_p)/(m_p+m_c)$.
The $p$-wave scattering states are those constructed in
Sec.~\ref{sec:scattering}, with the normalization fixed by the
Coulomb-matching convention of Eq.~\eqref{eq:scatt_asymptotic}.
For the entrance channels of interest ($p_{1/2}$ and $p_{3/2}$), the
calculated phase shifts remain small over the low-energy region
considered here.  Thus, the continuum is Coulomb dominated while still
being treated with the full Coulomb-matched scattering form in the
matrix elements.

\begin{figure*}[htpb]
\includegraphics[width=0.95\textwidth]{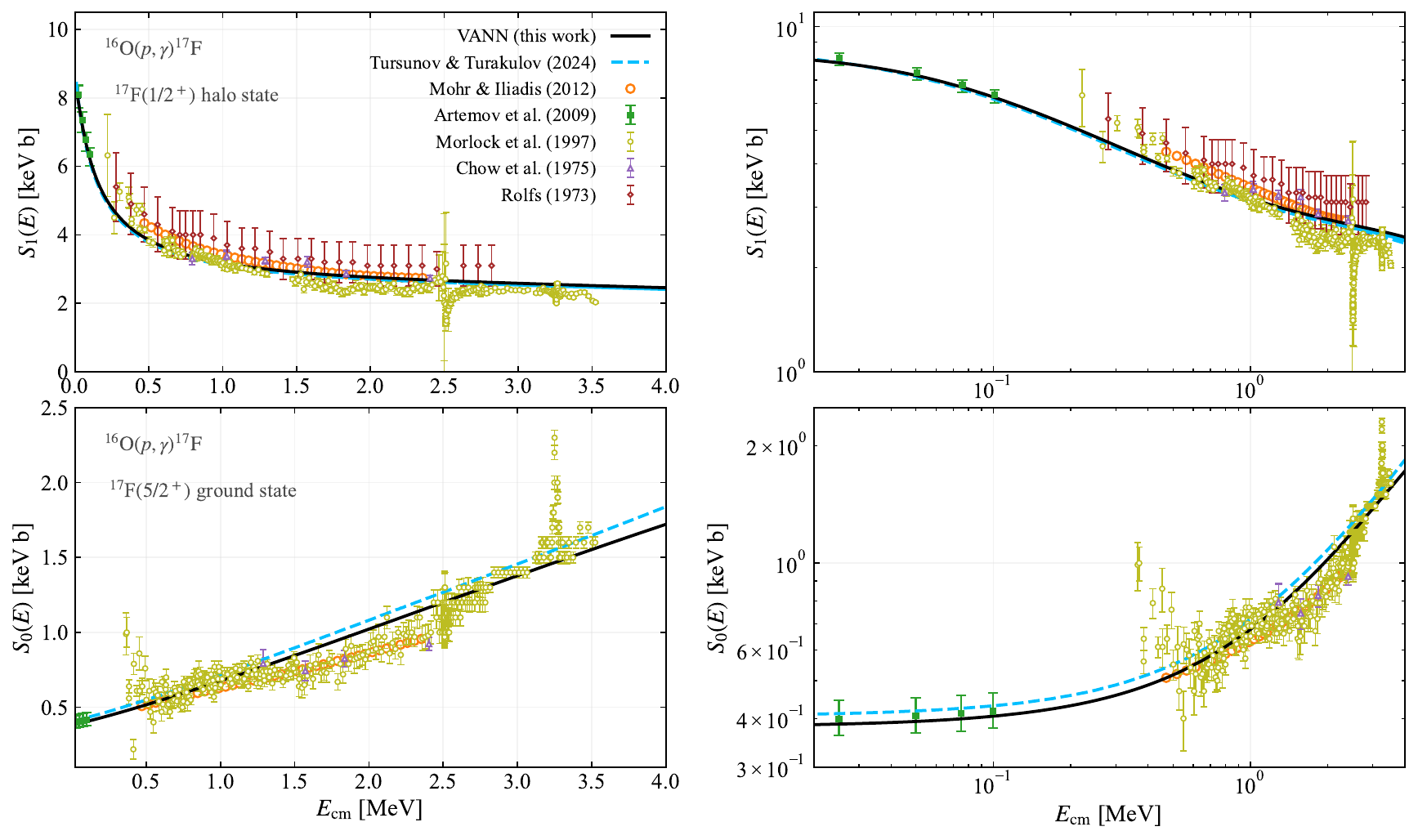}
\caption{Astrophysical $S$ factors for
$^{16}{\mathrm{O}}(p,\gamma)^{17}{\mathrm{F}}$.
Top panels: capture to the $s_{1/2}$ halo state.
Bottom panels: capture to the $d_{5/2}$ ground state.
The solid curves are the present VANN calculation using the hybrid
bound-state ansatz of Sec.~\ref{sec:ansatz}.
The dashed curves are the Tursunov--Turakulov benchmark for the same VM2
capture model~\cite{Tursunov2024}; this is a within-model validation.
Experimental data are from Mohr~\&~Iliadis~\cite{Mohr2012},
Artemov~\textit{et~al.}~\cite{Artemov2009},
Morlock~\textit{et~al.}~\cite{Morlock1997},
Chow~\textit{et~al.}~\cite{Chow1975},
and Rolfs~\cite{Rolfs1973}.}
\label{fig:sfactor}
\end{figure*}

The radial matrix element for a transition from an initial scattering
channel $\alpha\equiv(\ell_i j_i)$ to a final bound state $f$ is
\begin{equation}
I_\alpha(E)=\int_0^\infty u_f^{(b)}(r)\,r\,u_i^{(+)}(E,r)\,dr,
\label{eq:e1_integral}
\end{equation}
and the full $E1$ cross section sums the included channels incoherently,
\begin{equation}
\sigma(E)=\sum_\alpha \frac{2J_f+1}{2}
\frac{64\pi^2}{3\hbar}\frac{k_\gamma^3}{k^2v}
{\cal A}_\alpha e_{\mathrm{eff}}^2 |I_\alpha(E)|^2,
\end{equation}
where $k_\gamma=E_\gamma/(\hbar c)$,
$E_\gamma=E+S_p^{(f)}$, and $S_p^{(f)}=|E_f|$ is the proton separation
energy of final state $f$. The channel factor ${\cal A}_\alpha$ contains the
angular-momentum coupling and the statistical weight. For an $E1$ transition
from an initial scattering channel $\alpha\equiv(\ell_i j_i)$ to a final bound
state with orbital $\ell_f$ and total $J_f$, the reduced matrix element
angular factor is
\begin{equation}
\begin{split}
\left \langle (\ell_f\,\tfrac12)J_f||Y_1||(\ell_i\,\tfrac12)j_i \right \rangle
&=(-1)^{j_i+\ell_f+3/2}\,
\sqrt{[j_i][J_f]} \\
&\quad\times
\begin{Bmatrix}\ell_f&J_f&\tfrac12\\j_i&\ell_i&1\end{Bmatrix}
\langle 1\,0,\ell_i0|\ell_f0\rangle,
\end{split}
\label{eq:red_mat_e1}
\end{equation}
With $[x]=2x+1$, Eq.~\eqref{eq:red_mat_e1} gives squared reduced angular factors
$2/9$ and $4/9$ for the $p_{1/2}\to s_{1/2}$ and
$p_{3/2}\to s_{1/2}$ transitions, respectively, $4/5$ for
$p_{3/2}\to d_{5/2}$, and zero for $p_{1/2}\to d_{5/2}$.  Matching the
angular part of the source expression to the cross-section convention above
gives
\begin{equation}
 {\cal A}_\alpha=\frac{3|R_{fi}|^2}{4\pi[J_f]}.
\end{equation}
Thus no additional average over the entrance-channel total angular momentum
$j_i$ is introduced.  The resulting channel factors are listed in
Table~\ref{tab:channel_factors} and are the factors used to recompute
Figs.~\ref{fig:d52_tail_impact} and~\ref{fig:sfactor}.

\begin{table}[ht]
\centering
\caption{Channel factors ${\cal A}_\alpha$ for the $E1$ capture transitions
included in this work.}
\label{tab:channel_factors}
\setlength{\tabcolsep}{8pt}
\begin{tabular}{l c c c}
\toprule
Transition & $\ell_i$ & $j_i$ & ${\cal A}_\alpha$ \\
\midrule
$p_{1/2}\to s_{1/2}$ (halo)  & 1 & $1/2$ & $1/(12\pi)\approx0.02653$ \\
$p_{3/2}\to s_{1/2}$ (halo)  & 1 & $3/2$ & $1/(6\pi)\approx0.05305$ \\
$p_{3/2}\to d_{5/2}$ (g.s.)  & 1 & $3/2$ & $1/(10\pi)\approx0.03183$ \\
$p_{1/2}\to d_{5/2}$ (g.s.)  & 1 & $1/2$ & $0$ (forbidden) \\
\bottomrule
\end{tabular}
\end{table}

The astrophysical $S$ factor follows from
\begin{equation}
S(E)=E\,\sigma(E)\,\exp(2\pi\eta).
\label{eq:sfactor}
\end{equation}

In the present calculation, the bound-state input in Eq.~\eqref{eq:e1_integral}
is the trained VANN wave function. The continuum scattering states entering
the same equation are obtained independently by Numerov integration with
Coulomb matching, as described in Sec.~\ref{sec:scattering}. Numerov bound-state
solutions are used only as benchmarks, while the finite-difference
diagonalization is used independently to validate the Pauli-forbidden state.
It is worth reiterating that no target wave function,
target ANC, target radius, target node position, or empirical $S$-factor normalization is
used during VANN training.

Figure~\ref{fig:d52_tail_impact} quantifies the impact of the
asymptotic tail on the capture observables.  The exponential-envelope
control has nearly the same binding energy and a very similar interior
profile, but it underestimates the low-energy matrix element because its
large-distance amplitude is wrong.  At $E_{\mathrm{cm}}=20$~keV the resulting
$S$ factor differs by roughly a factor of $8.6$ from the Whittaker-tail
calculation, confirming that the tail error seen in Fig.~\ref{fig:d52_wave}
propagates directly into the capture observable.

Figure~\ref{fig:sfactor} shows the resulting $S$ factors.
The VANN calculation follows the Tursunov--Turakulov benchmark and the
experimental data at the level of tens of percent over most of the
plotted range, without any empirical rescaling of the VANN bound-state
wave function.
The ground-state branch shows comparable agreement.
The remaining differences with experimental compilations are not
interpreted as a failure of the VANN itself, but as the expected
limitation of the fixed two-body VM2 potential and of the
normalization conventions of the adopted capture model.
At 20~keV the VANN and Numerov calculations give, respectively,
$S_{\rm halo}=7.992$ and $8.043$~keV\,b; the corresponding ground-state
contributions are $S_0=0.386$ and $0.387$~keV\,b, and the totals are
$8.378$ and $8.430$~keV\,b.

\begin{figure}[htpb]
\includegraphics[width=0.95\columnwidth]{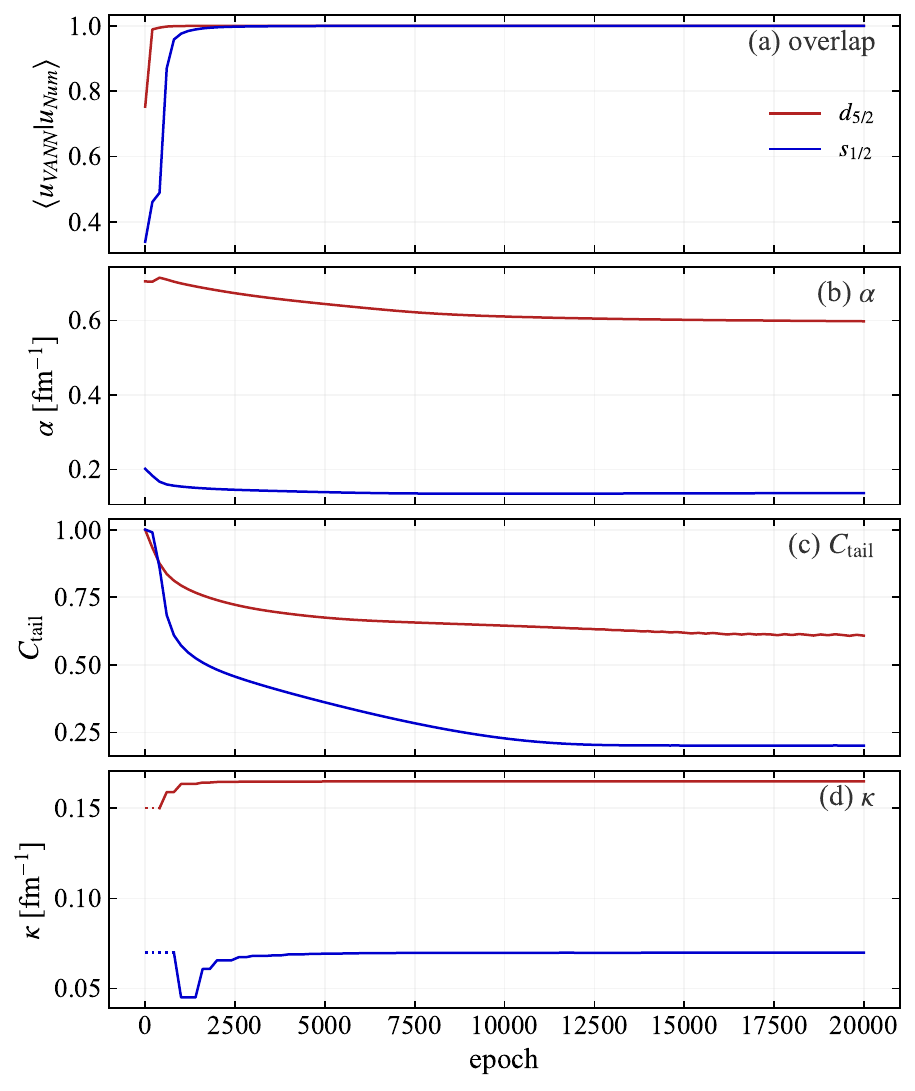}
\caption{Convergence histories from the production VANN runs used for the
bound-state results reported in Tables~\ref{tab:vann_params} and
\ref{tab:bound_summary}. The panels show: (a) the overlap with the corresponding
Numerov solution, (b) the trainable interior-envelope parameter $\alpha$,
(c) the Whittaker-tail amplitude $C_{\rm tail}$, and (d) the binding
momentum $\kappa_\theta$ used in the Whittaker tail.}
\label{fig:convergence_params}
\end{figure}
\section{Conclusions}
\label{sec:conclusions}
We have developed a variational neural-network representation for the
radial Schr\"odinger problem of $^{17}$F in a fixed two-body
$^{16}$O$+p$ Hamiltonian.  The calculation uses the V$_{\mathrm{M2}}$
interaction as a controlled test case in which the neural wave function
can be benchmarked against an independent Numerov solution.  Within this
setting, the VANN reproduces the compact $d_{5/2}$ ground state and the
weakly bound $s_{1/2}$ halo state in energy, node structure, rms radius,
and overlap, with the halo ANC differing by about $4.2\%$. The $s_{1/2}$ calculation first obtains the
deeply bound forbidden state by unrestricted VANN minimization and uses its
fixed normalized wave function in the Pauli penalty. Independent finite
differences validate that forbidden state. The selected one-node halo is
nearly orthogonal to the VANN forbidden reference, with overlap amplitude
$1.61\times10^{-3}$; its overlap with the independent FD state is
$5.36\times10^{-3}$.

The calculation demonstrates a crucial diagnostic: for peripheral capture, variational convergence 
of the energy is not sufficient to guarantee the correct asymptotic tail. A neural ansatz that matches 
the energy and interior can still fail dramatically in the ANC and $S$ factor unless the Coulomb–Whittaker 
asymptotic form is explicitly enforced. With the tail constrained, the VANN bound states reproduce the 
benchmark capture observables without any Numerov or empirical input during training.

For the extended halo, the residual ANC difference reflects the known
first-order sensitivity of asymptotic observables and is amplified by the
small binding-energy mismatch; it should not be interpreted as a
neural-network-specific failure. This contrasts with the second-order
stationarity of the energy and explains why a global overlap near unity is a
weaker test of a low-norm tail.

More broadly, the work shows how asymptotic boundary conditions can be
systematically embedded in a differentiable neural wave function, while the
Pauli-forbidden component is suppressed by an overlap penalty. Extensions to multi-channel
reaction models, parameter scans, inverse reconstructions, and three-body
calculations with coupled asymptotic constraints are natural future
directions, though they go beyond the two-body benchmark demonstrated
here.

\section*{Acknowledgments}

LAS thanks Prof. A. S. Tursunov and S. K. Nurmamatov for kindly
providing the numerical and experimental data corresponding to the Fig.~\ref{fig:sfactor}. 
TF acknowledges partial support from CNPq (Grant No. 306834/2022-7),  
INCT-FNA (Grant No. 464898/2014-5) and FAPESP (Grant No. 2023/13749-1 and 2025/05312-8).

\section*{Code and data availability}
The source code, input files, data, and scripts used to
generate the figures and tables are available from the corresponding
author upon reasonable request.

\section*{Appendix: Training parameters}

In addition to the physical observables shown in
Fig.~\ref{fig:convergence_physics}, the VANN training also tracks the auxiliary quantities that characterize the ansatz.
The overlap $\langle u_{\mathrm{VANN}}|u_{\mathrm{Num}}\rangle$ measures the
similarity between the VANN and Numerov;
it approaches unity as the two solutions become identical.
The parameter $\alpha$ controls the exponential decay of the interior
component $u_{\mathrm{int}}(r)=r^{\ell+1}e^{-\alpha r}{\mathrm{MLP}}(\ln(1+r))$;
it is trainable and is optimized together with the MLP weights.
The parameter $C_{\mathrm{tail}}$ sets the amplitude of the Whittaker tail
in Eq.~\eqref{eq:tail}.
The binding momentum $\kappa$ entering the same tail is updated from the
current negative Rayleigh energy according to Eq.~\eqref{eq:kappa}, rather
than optimized as an independent network parameter. As shown in
Fig.~\ref{fig:convergence_params}, these quantities stabilize during
training.

\bibliographystyle{apsrev4-2}
\bibliography{references}

@article{Tursunov2024, 
title = {Study of the direct 16O(p,γ)17F astrophysical capture reaction within a potential model approach},
journal = {Nuclear Physics A},
volume = {1051},
pages = {122931},
year = {2024},
issn = {0375-9474},
doi = {https://doi.org/10.1016/j.nuclphysa.2024.122931},
url = {https://www.sciencedirect.com/science/article/pii/S0375947424001131},
author = {E.M. Tursunov and S.A. Turakulov},
}

@article{Morlock1997,
  author  = {R. Morlock and others},
  title   = {Halo Properties of the First $1/2^+$ State in $^{17}{\rm F}$
             from the $^{16}{\rm O}(p,\gamma)^{17}{\rm F}$ Reaction},
  journal = {Phys.\ Rev.\ Lett.},
  volume  = {79},
  pages   = {3837},
  year    = {1997},
  doi     = {10.1103/PhysRevLett.79.3837},
}

@article{Borge1993,
  author  = {M. J. G. Borge and others},
  title   = {Beta-decay to the proton halo state in $^{17}{\rm F}$},
  journal = {Phys.\ Lett.\ B},
  volume  = {317},
  pages   = {25},
  year    = {1993},
  doi     = {10.1016/0370-2693(93)91564-4},
}

@article{KeebleRios2020,
  title = {Machine learning the deuteron},
  author = {Keeble, J. W. T. and Rios, A.},
  journal = {Physics Letters B},
  volume = {809},
  pages = {135743},
  year = {2020},
  doi = {10.1016/j.physletb.2020.135743}, 
}

@article{Chow1975,
  author  = {H. C. Chow and G. M. Griffiths and T. H. Hall},
  title   = {The $^{16}{\rm O}(p,\gamma)^{17}{\rm F}$ Direct Capture
             Cross Section with an Extrapolation to Astrophysical Energies},
  journal = {Can.\ J.\ Phys.},
  volume  = {53},
  pages   = {1672},
  year    = {1975},
  doi     = {10.1139/p75-213},
}

@article{Rolfs1973,
  author  = {C. Rolfs},
  title   = {Spectroscopic factors from radiative capture reactions},
  journal = {Nucl.\ Phys.\ A},
  volume  = {217},
  pages   = {29},
  year    = {1973},
  doi     = {10.1016/0375-9474(73)90622-2},
}

@article{Ryberg2013,
  author  = {E. Ryberg and C. Forss{\'e}n and H.-W. Hammer and L. Platter},
  title   = {Effective field theory for proton halo nuclei},
  journal = {Phys. Rev. C},
  volume  = {89},
  pages   = {014325},
  year    = {2014},
  doi     = {10.1103/PhysRevC.89.014325}, 
}

@article{Hagen2010,
  author  = {G. Hagen and T. Papenbrock and M. Hjorth-Jensen},
  title   = {{\it Ab-initio} computation of the $^{17}$F proton-halo state
             and resonances in $A=17$ nuclei},
  journal = {Phys.\ Rev.\ Lett.},
  volume  = {104},
  pages   = {182501},
  year    = {2010},
  doi     = {10.1103/PhysRevLett.104.182501},
}

@article{Du2026,
  author  = {X.-K. Du and S.-Q. Zhang},
  title   = {Solving nuclear Dirac Woods--Saxon potential with a
             physics-informed neural network},
  journal = {Commun.\ Theor.\ Phys.},
  volume  = {78},
  number  = {3},
  pages   = {035303},
  year    = {2026},
  doi     = {10.1088/1572-9494/ae1a5c},
}

@article{Blue1965,
  author  = {R. A. Blue and W. Haeberli},
  title   = {Elastic scattering of protons by $^{16}$O},
  journal = {Phys.\ Rev.},
  volume  = {137},
  pages   = {B284},
  year    = {1965},
  doi     = {10.1103/PhysRev.137.B284},
}

@article{Bennaceur2000,
  author  = {K. Bennaceur and others},
  title   = {Shell model description of $^{16}{\rm O}(p,\gamma)^{17}{\rm F}$
             and $^{16}{\rm O}(p,p)^{16}{\rm O}$ reactions},
  journal = {Phys.\ Lett.\ B},
  volume  = {488},
  pages   = {75},
  year    = {2000},
  doi     = {10.1016/S0370-2693(00)00843-1},
}

@article{karniadakis2021piml,
  title = {Physics-informed machine learning},
  author = {Karniadakis, George Em and Kevrekidis, Ioannis G. and Lu, Lu and Perdikaris, Paris and Wang, Sifan and Yang, Liu},
  journal = {Nature Reviews Physics},
  volume = {3},
  pages = {422--440},
  year = {2021},
  doi = {10.1038/s42254-021-00314-5}
}

@article{carleo2017nqs,
  title = {Solving the quantum many-body problem with artificial neural networks},
  author = {Carleo, Giuseppe and Troyer, Matthias},
  journal = {Science},
  volume = {355},
  number = {6325},
  pages = {602--606},
  year = {2017},
  doi = {10.1126/science.aag2302}
}

@article{LeCun2015,
  title = {Deep learning},
  author = {LeCun, Yann and Bengio, Yoshua and Hinton, Geoffrey},
  journal = {Nature},
  volume = {521},
  pages = {436--444},
  year = {2015},
  doi = {10.1038/nature14539}
}

@article{Lei2026,
  author  = {Lei, Jin},
  title   = {Exterior complex scaling enables physics-informed neural networks for nuclear reactions},
  journal = {Phys. Rev. C},
  volume  = {113},
  pages   = {064618},
  year    = {2026},
  doi     = {10.1103/sjz4-pq6p}
}

@article{raissi2019pinns,
  title = {Physics-informed neural networks: A deep learning framework for solving forward and inverse problems involving nonlinear partial differential equations},
  author = {Raissi, Maziar and Perdikaris, Paris and Karniadakis, George Em},
  journal = {Journal of Computational Physics},
  volume = {378},
  pages = {686--707},
  year = {2019},
  doi = {10.1016/j.jcp.2018.10.045}
}

@article{Mohr2012,
title = {Recommended cross-section of the $^{16}$O$(p,\gamma)^{17}$F reaction below 2.5 MeV: A potential tool for quantitative analysis and depth profiling of oxygen},
author = {Mohr, P. and Iliadis, C.},
journal = {Nucl. Instrum. Methods Phys. Res. A},
volume = {688},
pages = {62--65},
year = {2012},
doi = {10.1016/j.nima.2012.05.084}
}

@article{Artemov2009,
  author  = {Artemov, S. V. and Igamov, S. B. and Tursunmakhatov, K. I. and Yarmukhamedov, R.},
  title   = {Determination of nuclear vertex constants (asymptotic normalization coefficients) for the virtual decays {${}^3$He} $\rightarrow$ d + p and {${}^{17}$F} $\rightarrow$ {${}^{16}$O} + p and their use for extrapolating astrophysical {S}-factors of the radiative proton capture by the deuteron and the {${}^{16}$O} nucleus at very low energies},
  journal = {Bulletin of the Russian Academy of Sciences: Physics},
  volume  = {73},
  number  = {2},
  pages   = {165--170},
  year    = {2009},
  month   = feb,
  doi     = {10.3103/S1062873809020075},
  url     = {https://doi.org/10.3103/S1062873809020075},
  issn    = {1934-9432}
}

@article{Iliadis2022,
  title = {Bayesian estimation of the $S$ factor and thermonuclear reaction rate for $^{16}\mathrm{O}(p,\ensuremath{\gamma})^{17}\mathrm{F}$},
  author = {Iliadis, Christian and Palanivelrajan, Vimal and de Souza, Rafael S.},
  journal = {Phys. Rev. C},
  volume = {106},
  issue = {5},
  pages = {055802},
  numpages = {11},
  year = {2022},
  month = {Nov},
  publisher = {American Physical Society},
  doi = {10.1103/PhysRevC.106.055802},
  url = {https://link.aps.org/doi/10.1103/PhysRevC.106.055802}
}

@article{LiLuoSunOrtega2026,
    author = "Li, Ruitian and Luo, Xuan and Sun, Hao and Ortega, Pablo G.",
    title = "{Solving two and three-body systems with deep neural networks}", 
    doi = "10.1140/epjc/s10052-026-15756-3",
    journal = "Eur. Phys. J. C",
    volume = "86",
    number = "5",
    pages = "503",
    year = "2026"
}

@article{Iliadis2008,
  title = {New reaction rate for $^{16}\mathrm{O}$($p,\ensuremath{\gamma}$)$^{17}\mathrm{F}$ and its influence on the oxygen isotopic ratios in massive AGB stars},
  author = {Iliadis, C. and Angulo, C. and Descouvemont, P. and Lugaro, M. and Mohr, P.},
  journal = {Phys. Rev. C},
  volume = {77},
  issue = {4},
  pages = {045802},
  numpages = {11},
  year = {2008},
  month = {Apr},
  publisher = {American Physical Society},
  doi = {10.1103/PhysRevC.77.045802},
  url = {https://link.aps.org/doi/10.1103/PhysRevC.77.045802}
}

@article{Rozalen2024,
  author  = {Rozal{\'e}n Sarmiento, J. and Keeble, J. W. T. and Rios, A.},
  title   = {Machine learning the deuteron: new architectures and uncertainty quantification},
  journal = {Eur. Phys. J. Plus},
  volume  = {139},
  pages   = {189},
  year    = {2024},
  doi     = {10.1140/epjp/s13360-024-04983-w}
}

@article{Saito1969,
  author  = {Saito, Sakae},
  title   = {Interaction between Clusters and Pauli Principle},
  journal = {Prog. Theor. Phys.},
  volume  = {41},
  pages   = {705--722},
  year    = {1969},
  doi     = {10.1143/PTP.41.705}
}

\end{document}